\documentclass[12pt,twoside]{article}
\usepackage{fleqn,espcrc1}
\usepackage{graphicx}

\def\raw{\rightarrow}

\def\be{\begin{equation}}
\def\ee{\end{equation}}
\def\bea{\begin{eqnarray}}
\def\eea{\end{eqnarray}}
\def\bear{\begin{array}}
\def\ear{\end{array}}
\def\bfig{\begin{figure}}
\def\efig{\end{figure}}
\def\bcen{\begin{center}}
\def\ecen{\end{center}}

\def\la{\label}
\def\chic{\scriptscriptstyle}

\def\Jl#1#2#3#4{{#1} {#2} (19#3) #4}
\def\NPA{Nucl. Phys. A}
\def\NPB{Nucl. Phys. B}
\def\PLB{Phys. Lett. B}

\def\PRC{Phys. Rev. C}
\def\PRD{Phys. Rev. D}

\def\ZPhA{Z. Phys. A}

\title{Charged current weak production of the $\Delta$ resonance}

\author{L. Alvarez-Ruso
\address{Departamento de F\'{\i}sica Te\'{o}rica and IFIC, Centro Mixto
Universidad de Valencia-CSIC, 46100 Burjassot, Valencia, Spain}
, E. Oset~$^\mathrm{a}$  
, S. K. Singh 
\address{Physics Department, Aligarh Muslim University, 202002 Aligarh, India}
, M. J. Vicente Vacas~$^\mathrm{a}$}

\begin{document}
\maketitle
\begin{abstract}
The reactions $e^{-}\,p \rightarrow \Delta^{0}\, \nu_{e}$ and
$e^{+}\,p \rightarrow \Delta^{++}\, \bar{\nu}_{e}$  are considered as a
possible source of information about the weak $N\Delta$ transition form
factors. The low $q^2$ BNL data on $\nu_\mu$ production of $\Delta$ are used
to extract the axial vector $N\Delta$ coupling, taking into account the
deuteron structure and the $\Delta$ width. Finally, pion production induced by
neutrinos in $^{16}O$ in the $\Delta$ region, relevant to atmospheric $\nu$
experiments, is investigated. 
\end{abstract}

\section{INTRODUCTION}

The nucleon excitation spectrum is a valuable source of information about
baryon structure. 
The $N\Delta$ transition presents clear advantages from the
experimental point of view since the $\Delta$ is separated from the rest
of resonances. 
The bulk of the existing information on the weak $N\Delta$ transition form
factors (FF) comes from the analysis of the ANL \cite{rade} and BNL \cite{kita}
experiments, performed with $\nu_\mu$ beams, whose energies span from 0.5 to
6.0~GeV with poorly known distributions.
Nowadays, with the advent of the
new generation of electron accelerators in the GeV region and
achieving high luminosities, it is possible to perform electron scattering
experiments in the resonance region. We have considered the possibility to
extend these studies to the weak charged current physics. 
For this reason, we have studied the reactions
$e^{-}\,p \rightarrow \Delta^{0}\, \nu_{e}$ and
$e^{+}\,p \rightarrow \Delta^{++}\, \bar{\nu}_{e}$ at the typical energies
of MAMI and TJNAF, and using the available information about the FF~\cite{yo3}.

Since the vector $N\Delta$ FF are related to the isovector
electromagnetic ones, which can be obtained from electroproduction data,
these experiments would allow to study the axial FF and, in
particular, the dominant $C_5^A$. The determination of its value at
$q^2=0$ is important in view of the discrepancies between the PCAC prediction 
and theoretical estimates obtained in most quark models \cite{mukho2}.
We have used the low $q^2$ BNL data on the ratio of $\mu^- \Delta^{++}$
and $\mu^- p$ events from $\nu_\mu d$ collisions to extract the value of the
axial vector coupling $C_5^A(0)$, taking into account the deuteron structure
and the $\Delta$ width~\cite{yo4}. 

The study of weak $N\Delta$ transitions in nuclei is relevant for
the analysis of atmospheric neutrino experiments. In fact, the energy
distribution of the part of the atmospheric $\nu$ flux producing fully
contained events at Kamiokande is such that $< E_\nu > \approx 700$~MeV,
well above the $\Delta$ production threshold. These $\Delta$'s decay
into pions and photons (through $\pi^0$ decay), that are a source of
background. For this reason, we have studied the impact of nuclear effects in
$\nu_{e(\mu)}$ production of $\Delta$ in $^{16}O$ \cite{nucl}.

\section{WEAK ELECTROPRODUCTION CROSS SECTION}

The matrix element for the process 
$e^{-}(k)+p(p) \rightarrow \Delta ^{0}(p')+ \nu_{e}(k')$
is proportional to the product of the leptonic and hadronic currents.
The hadronic current is
expressed in terms of vector and axial vector FF
$C_i^V$ and $C_i^A$ ($i=3,4,5,6$) \cite{yo3}. The
imposition of the CVC hypothesis $q_\mu J_V^\mu = 0$ implies $C_6^V=0$.
The other three vector FF are obtained from the isovector
electromagnetic ones. Assuming $M1$ dominance, one gets
$C_5^V=0$ and $C_4^V=-\left(M / M_\Delta\right) C_3^V$.
$C_3^V$ is determined from electroproduction experiments
\cite{hipp} and from a quark model \cite{liu}
\begin{eqnarray}
\la{set1v}
C_3^V(q^2) & = & 2.05 \,(1-q^2/0.54\,{\rm GeV}^2)^{-2}\,, \\[0.2cm]
\la{set3v}
C_3^V(q^2) & = & M / (\sqrt{3}m)\, e^{-{\bar q}^2/6} \,,
\end{eqnarray}
where $m = 330$~MeV is the quark mass and
$\bar q= |\mathbf{q}|/ \alpha_{\chic{HO}}$, with $\alpha_{\chic{HO}}=320$~MeV.
Concerning the axial FF, $C_6^{A}$ can be related to $C_5^{A}$ using pion
pole dominance and PCAC, then
$C_6^A(q^2) = C_5^A(q^2) M^2/\left(m_{\pi}^2-q^2\right)$.
The value of $C_5^A(0)$ can be taken from the off-diagonal
Goldberger-Treiman relation \cite{hipp}, 
$C_5^A(0) = g_{\Delta N \pi} f_{\pi} / (\sqrt{6} M) = 1.15$, 
where $f_\pi = 92.4$~MeV, $g_{\Delta N \pi} = 28.6$; $C_3^A(q^2)$, 
$C_4^A(q^2)$ and $C_5^A(q^2)/C_5^A(0)$ are given by the Adler model
\cite{adler}
\begin{equation}
\la{set12}
C_{i=3,4,5}^A(q^2) = {{C_i(0)\left[ 1-{{a_i q^2}\over{b_i-q^2}} \right] }
{\left( 1- {{q^2}\over{M_A^2}}\right)^{-2}}}\,.
\end{equation}
with $C_3^A(0)=0$, $C_4^A(0)=-0.3$, $a_4=a_5=-1.21$,
$b_4=b_5=2$~GeV$^2$ and $M_A=1.28$~GeV.
The value of $M_A$ comes from a best fit to the $\mu^- \Delta^{++}$ events 
at BNL \cite{kita}. For a comparison, we also use a
non-relativistic quark model calculation \cite{liu}
\begin{equation}
\la{set3}
C_5^A(q^2)=\left(\frac{2}{\sqrt{3}}+\frac{1}{3\sqrt{3}}\frac{q_0}{m}\right) 
e^{-{\bar q}^2/6} \ , \ C_4^A(q^2)=-\frac{1}{3\sqrt{3}}\frac{M^2}{M_\Delta m}
e^{-{\bar q}^2/6} \ , \ C_3^A(q^2)=0.
\end{equation}

From the amplitude given above, the differential cross section
$d \sigma/ d \Omega_{\Delta}$ can be obtained in the standard way.
The $\Delta$ width has been accounted for by means of the substitution
\be 
\delta(p'^2 - M_\Delta^2)\, \raw \, -\frac{1}{\pi}\frac{1}{2 M_\Delta}
\mathrm{Im}\left[ \frac{1}{W - M_\Delta + \frac{1}{2}i\Gamma_\Delta} 
\right],\,
\Gamma_\Delta= \Gamma_0 \frac{M_\Delta}{W} 
\frac{q^3_{c.m.}(W)}{q^3_{c.m.}(M_\Delta)}\,,\,W= \sqrt{p'^2}
\ee
with $q_{c.m.}$ being the pion momentum in the $\Delta$ rest frame and
$\Gamma_0 = 120$~MeV. The angular distribution is shown in Fig.~\ref{fig1}
\begin{figure}[h!]
\begin{center}
\includegraphics[height=0.8\textwidth, angle=-90]{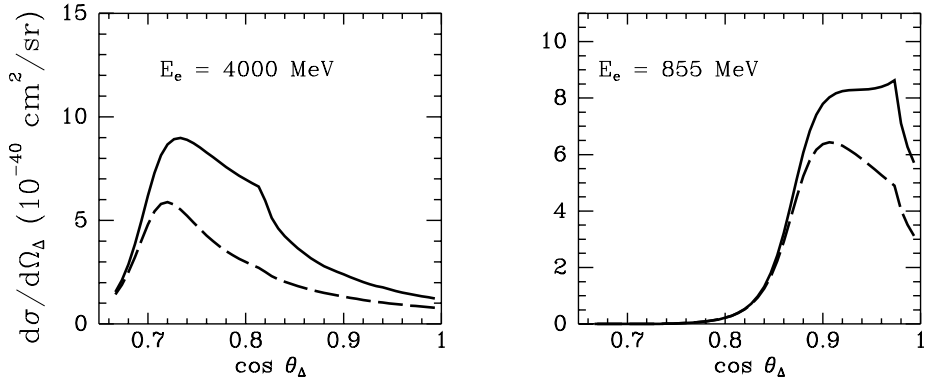}
\caption{$\Delta^0$ angular distribution in 
$e^{-}(k)+p(p) \rightarrow \Delta ^{0}(p')+ \nu_{e}(k')$.}
\la{fig1}
\end{center}   
\end{figure}
for two different sets of FF: I, phenomenological [Eqs.~(\ref{set1v}),
(\ref{set12})], solid line; 
II, quark model, [Eqs.~(\ref{set3v}), (\ref{set3})], dashed line. The
invariant mass has been restricted to $W < 1.4$~GeV to select
$\Delta$ events. The differential cross section is found to
be high enough in a large angular region to consider the possibility of
measuring them. 

\section{DETERMINATION OF THE AXIAL VECTOR COUPLING}

In order to obtain $C_5^A(0)$ we have evaluated the ratio 
\begin{equation}
\la{ra1}
R(Q^2)=\frac{\left(d \sigma/d q^2\right) \left(\nu d \rightarrow 
\mu^- \Delta^{++} n\right)}{\left(d \sigma/d q^2\right)\left(\nu d \rightarrow 
\mu^- p p\right)}\,, \quad Q^2 = -q^2
\end{equation}
at $E_\nu = 1.6$~GeV, which is the mean energy of the BNL $\nu_\mu$ spectrum;
the $\Delta$ production cross section has been calculated in the
impulse approximation, and using the deuteron wave function of the Paris
potential. 
The quasielastic cross section, in the same approximation, is taken from
Ref.~\cite{singh}. We found that, in the data region i.e. at
$Q^2 \leq 0.1$~GeV$^2$, deuteron effects are negligible and, hence, one
can treat the BNL data as if they were data on the ratio of the free reactions
\begin{equation}
\la{ra2}
R(Q^2) \approx R_0(Q^2)=\frac{\left(d \sigma/d q^2\right)\left(\nu p \rightarrow 
\mu^- \Delta^{++} \right)}{\left(d \sigma/d q^2\right)\left(\nu n \rightarrow 
\mu^- p \right)}\,.
\end{equation}
At $Q^2=0$, $R_0(Q^2)$ is given by the quotient of 
\be
\la{delta0}
\frac{d \sigma}{d q^2} =  \left({C^{A}_5}\right)^2 
\, \frac{1}{24 \pi^2}
G^2 \cos^2\theta_c \frac{\sqrt{s} (M+M_\Delta)^2 (s-M_\Delta^2)^2}{(s-M^2) 
M_\Delta^3} \int 
d k'^0 \frac{\Gamma_\Delta(W)}{(W-M_\Delta)^2+\Gamma_\Delta^2(W)/4}
\ee
and the well known expression for the forward quasielastic cross section.
Equating this ratio to the experimental value $0.55 \pm 0.05$ \cite{rade},
we obtain $C^{A}_5 = 1.22 \pm 0.06$; this result is consistent with the
value given by the off-diagonal Goldberger-Treiman relation. The proper
inclusion of the $\Delta$ width causes a 30~\% reduction of the cross
section and cannot be neglected in the extraction of $C^{A}_5(0)$.

\section{NEUTRINO PRODUCTION OF $\mathbf{\Delta}$ IN $\mathbf{^{16}O}$}

When the reactions $\nu_l\, p(n) \raw l^-\, \Delta^{++}(\Delta^{+})$
and $\bar{\nu}_l\, p(n) \raw l^+\, \Delta^{0}(\Delta^{-})$
take place
in the nucleus, the nucleon momentum is constrained within a density
dependent Fermi sea. The produced $\Delta$ does not have this
constraint, but its decay is inhibited by the Pauli blocking of the final
nucleon. On the other side, there are other disappearance channels open
through particle-hole excitations. The situation is well described if one
replaces in the $\Delta$ propagator
$\Gamma_\Delta \raw \hat{\Gamma}_\Delta - 2 \mathrm{Im}\Sigma_\Delta$ and
$M_\Delta \raw M_\Delta + \mathrm{Re}\Sigma_\Delta$,  
where $\hat{\Gamma}_\Delta$ is the Pauli blocked decay width and 
$\Sigma_\Delta$ is the $\Delta$ selfenergy in the nuclear medium \cite{oset1}.
The pions produced inside
the nucleus are rescattered and absorbed in their propagation through the
nucleus. The absorption coefficient required to estimate the produced pion
flux has been calculated in the eikonal approximation, taking the pion
energy dependent mean free path from Ref.~\cite{oset2}. 
For the $N\Delta$ transition FF, the phenomenological set I described above
has been taken; possible medium modification of the FF has not been 
considered.

In Fig.~\ref{fig2}~a) $d \sigma/d E_{k'}$ ($k'$ being the momentum of the
outgoing electron) is shown for $E_\nu = 750$~MeV. The medium modification
effects cause an overall reduction of about 40~\%. Therefore, the
Kamiokande analysis, which makes use of free $\Delta$ production cross
sections, overestimates one pion production. However, as can be seen in
Fig.~\ref{fig2}~b), the ratio of total pion production cross sections 
induced by electron and muon type neutrinos and antineutrinos
$R(E_\nu) = \sigma_\Delta(\mu)/\sigma_\Delta(e)$ is not affected by these
modifications.
\begin{figure}[h!]
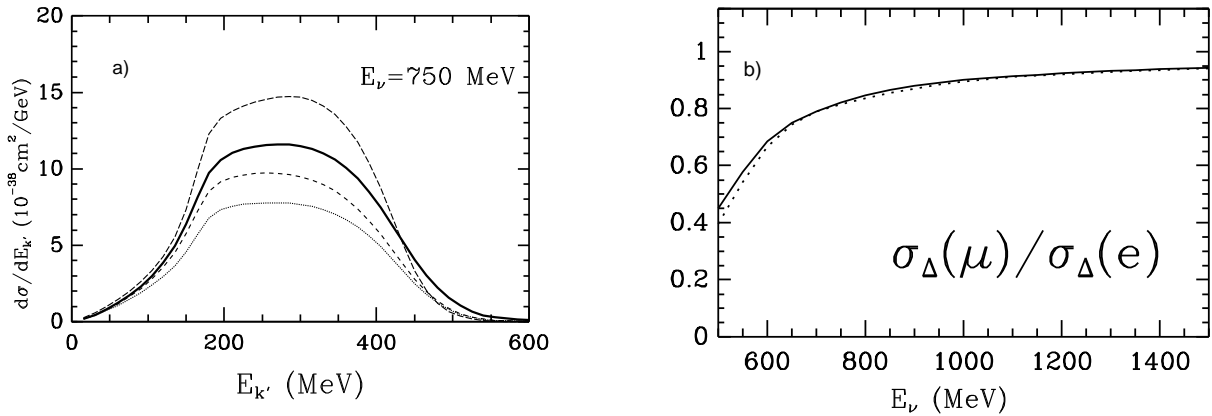

\begin{minipage}{.45\linewidth}
\begin{center}
\includegraphics[width=\linewidth]{delta2.ps}
\end{center}
\end{minipage}
\hfill
\begin{minipage}{.45\linewidth}
\begin{center}
\includegraphics[width=\linewidth]{delta3.ps}
\end{center}
\end{minipage}
\begin{center}
\caption{(a) $\nu_e$ induced $\Delta$ excitation in $^{16}O$ without 
(long-dashed line) and with medium effects (solid line); pion production
with medium effects, without (short-dashed line) and with absorption (dotted
line). (b) $R(E_\nu) = \sigma_\Delta(\mu)/\sigma_\Delta(e)$ with (solid
line) and without medium effects (dotted line).} 
\la{fig2}
\end{center}
\end{figure}
\vspace{-1.5cm}

\section{ACKNOWLEDGEMENTS}

L.A.R. acknowledges financial support from the Generalitat Valenciana and 
S.K.S., from the Spanish Ministerio de Educaci\'on y Cultura. This work has
been partially supported by DGYCIT contract PB 96-0753.

\end{document}